\documentclass{llncs}

\usepackage{booktabs} 
\usepackage{hyperref}
\usepackage{multirow}

\usepackage[colorinlistoftodos]{todonotes} 

\graphicspath{ {images/} }

\begin{document}
\title{Complete Semantics to empower Touristic Service Providers\thanks{Completeness is something that can never be achieved. Therefore we think it is a proper goal to target our ambitions.}}

\author{Zaenal Akbar, Elias K\"{a}rle, Oleksandra Panasiuk, Umutcan \c{S}im\c{s}ek, Ioan Toma, Dieter Fensel}
\institute{Semantic Technology Institute (STI) Innsbruck, University of Innsbruck, Austria \email{firstname.lastname@sti2.at}}

\maketitle              

\begin{abstract}
The tourism industry has a significant impact on the world's economy, contributes 10.2\% of the world's gross domestic product in 2016. 
It becomes a very competitive industry, where having a strong online presence is an essential aspect for business success. To achieve this goal, the proper usage of latest Web technologies, particularly schema.org annotations is crucial. 
In this paper, we present our effort to improve the online visibility of touristic service providers in the region of Tyrol, Austria, by creating and deploying a substantial amount of semantic annotations according to schema.org, a widely used vocabulary for structured data on the Web.
We started our work from Tourismusverband (TVB) Mayrhofen-Hippach and all touristic service providers in the Mayrhofen-Hippach region and applied the same approach to other TVBs and regions, as well as other use cases.
The rationale for doing this is straightforward. Having schema.org annotations enables search engines to understand the content better, and provide better results for end users, as well as enables various intelligent applications to utilize them.
As a direct consequence, the region of Tyrol and its touristic service increase their online visibility and decrease the dependency on intermediaries, i.e. Online Travel Agency (OTA).
\end{abstract}

\keywords{Semantic annotations, schema.org, touristic service providers}

\section{Introduction}
\label{sec:Introduction}
The tourism and leisure industry contributes significantly to the economic development of the region of Tyrol, Austria.
With around 60,000 employees (25\% of the full-time workplaces in the region were created in this industry), it generates sales approximately 8.4 billion Euros.
In the tourism year 2015/2016, 11.5 million guests were arrived, generated 47.6 million overnight stays.
The direct value added of the industry to the region is 17.5\%, higher than other regions such as Upper Austria (3.1\%), Vienna (1.6\%), or national level (5.3\%)\footnote{Tirol Werbung, 2016, ``Der Tiroler Tourismus Zahlen, Daten und Fakten 2016'', \url{http://www.tirolwerbung.at/tiroler-tourismus/zahlen-und-fakten-zum-tiroler-tourismus/}}.

The TVB Mayrhofen-Hippach\footnote{\url{http://www.mayrhofen.at}} is the tourism board of the Mayrhofen-Hippach region situated in Zillertal, Tyrol, Austria.
It is the organization responsible for the marketing of the entire Mayrhofen-Hippach region and its members. As with all touristic service providers, it faces the challenge of achieving the highest visibility possible in search engines, and at the same time, they need to be present in various communication channels which are constantly growing.
Website, for example, bridges the tourism organizations and tourists directly and plays roles at different stages of tourists decision making process~\cite{Gupta2014}. Information quality, responsiveness, visual appearance, personalization are a few example of key factors for influencing website effectiveness.
Specifically for the Alps region, the regional tourism boards have been enhanced their websites qualities significantly in various dimensions, not only information quality but also the adoption of new technologies including a few web standards and interactive maps~\cite{Mich2013}.

But it is still challenging for the tourism sector, especially in the region of Tyrol, to provide useful content that could help potential guests to make a reservation decision directly as well as to be accessible by machine, i.e. semantically annotated~\cite{hepp2006}. 
In Austria national scope, most of the touristic service providers have not or minimally use the semantic annotations technology~\cite{Stavrakantonakis2013}.
And most of the existing annotations of touristic service providers, especially hotels, were performed incorrectly~\cite{karle2016there}. This situation is critical for the industry because the use of semantic annotation such as schema.org\footnote{\url{http://schema.org}} could increase a typical hotel website visibility by 20\%~\cite{Fensel2016}.
More than just for increasing online visibility on search engines, a semantically annotated content of touristic service providers within a region could contribute to the tourism information system of the region. For example, enabling data query from distributed sources, topical or location-based data integration, matching of service providers and requesters, as well as transactional web services for tourists~\cite{maedche2002}. 
And we believe that the seamless interoperability among organizations which is still an issue in the tourism industry~\cite{Werthner2015} can be solved by semantic web technologies including semantic annotations.

A substantial amount of semantically annotated content (possibly as complete as possible from every touristic service providers and touristic related information sources) could support every intelligent, machine processing decision making for the industry.
Search engines such as Google consume annotated content and present it in a more interesting way visually such as stars for ratings instead of text, a structured layout for events, carousels for recipes\footnote{\url{https://developers.google.com/search/docs/guides/search-features}}. With those richer search results, content annotation approach outperforms the conventional search engine optimization techniques.
Annotated content also could help organizations to semi-automatically disseminate content to multiple online communication channels~\cite{Akbar2014}, reducing human efforts to manually collecting, curating content from different sources before distributing them to multiple channels. Most recently, semantically annotated content will be consumed by intelligent applications such as chatbot and personal digital assistant to automatically provide users with precise and personalized information.

This paper describes our systematic approach on annotating tourism information available on the region of Tyrol, Austria, started with the TVB Mayrhofen-Hippach website using semantic annotations, more precisely schema.org. 
The main goal of our cooperation, therefore, is to improve the online visibility of the region by enriching their content with machine processable data.
Comparing to our similar efforts before, where we have been annotating various individual service providers such as hotel~\cite{Fensel2016}, this is by far the biggest effort regarding covered information sources, the number of produced annotation, types of annotations as well as how often an update need to be performed.
More precisely the contributions of this paper are as follows: 
(1) an approach to \textbf{automatically generate semantic annotations} of dynamic data based on data APIs, as well as manually generate semantic annotations for the static data\footnote{Dynamic data change rather frequently e.g. hotel offers and events, while static data change very rarely, e.g. contact information of a hotel.} 
(2) a method to \textbf{link the semantic annotations with the content} and 
(3) a regularly updated \textbf{schema.org annotations}~\footnote{Per March 29, 2017, we have generated 1,6 Million (1,567,254 to be precise) triples of annotations for TVB Mayrhofen-Hippach} generated using a mixed approach i.e. automatically as well as manually. 
The rest of this paper is organized as follows. Section~\ref{sec:RelatedWork} describes related approaches that aim to address the creation of semantic annotations at large scale. 
Section~\ref{sec:Methodology} describes our methodology for identifying information that needs to be annotated and what types of annotations will be provided. Section~\ref{sec:Implementation} shows our implementation on annotating the tourist board Mayrhofen-Hippach website with schema.org, including its current results.
Section~\ref{sec:OngoingWork} outlines our ongoing work on intelligently utilizing the obtained annotations, and finally, Section~\ref{sec:Conclusion} concludes our paper and describes our future work.

\section{Motivation and Related Work}
\label{sec:RelatedWork}
The development in the mobile computing and artificial intelligence is leading the way to the development of a new layer on top of the web, so-called ``headless web''\footnote{https://paul.kinlan.me/the-headless-web/}, where the presentation of the web pages loses significance and publishing semantically described structured data becomes more important than ever. Among many vocabularies for embedding semantic data into the webpages, schema.org comes to the fore as a de-facto standard. Schema.org offers set of vocabularies that facilitate the publication of structured data on the web and it has been evolved rapidly since its introduction in 2011. The success of schema.org can be measured by its adoption rate. The results of the Web Data Commons crawl in October 2016 shows that the web pages with triples are 39\% of the overall crawled web pages which is 8\% higher than the previous year\footnote{For detailed statistics: \url{http://webdatacommons.org/structureddata/}}. The increased support from Content Management Systems (CMS) for schema.org as well as the support from other third-party software for tasks like event management have a great impact on this wide adoption~\cite{Guha:2016:SES:2886013.2844544}.

When it comes to the tourism sector, the scene is quite different. Although the amount of schema.org annotations increase among the touristic service providers, the annotations usually come not directly from the lodging business' website (e.g. a hotel website), but from an entity like an online travel agency. Moreover, the annotations are usually incorrect or incomplete (e.g. missing values for important properties such as address)~\cite{karle2016there}. 
Even though CMS helps to publish a significant amount of structured content on the web, especially for the tourism sector, there is still a lot of data stored in the databases of proprietary software and served with an API. The publication of such data (e.g. events, offers from a hotel) described with schema.org carries a great importance in terms of online visibility and e-commerce in the headless web. Additionally, it will also contribute to their visibility on the search engines through features like Rich Snippets~\cite{Goel2009}.

Given the results of the aforementioned analysis, this endeavor is challenging for two reasons: (a) there is a big development effort required from various parties to generate and publish structured data based on the existing internal data, (b) the lack of know-how of the touristic service providers and software producers in tourism field in terms of mapping internal metadata to correct schema.org types and properties. To tackle this challenge, we provide a solution that requires minimal development effort and know-how for the touristic service providers. 

From a syntactic point of view, there are various ways to include schema.org into web pages, namely Microdata, RDFa, and JSON-LD. Microdata and RDFa have been around for many years and gained widespread usage~\cite{Bizer2013}. Unlike Microdata and RDFa, JSON-LD does not require the annotations to be directly embedded in HTML markup blocks where the content reside, but it can be placed anywhere in the source of the web page in script tags. This is one of the main reasons we adopt JSON-LD for our implementation since it brings an advantage for dynamic injection of the semantic annotation to the web pages. The annotations can be prepared and hosted externally and be embedded on demand straightforwardly.
There is a major effort from the semantic web and linked data community for generating semantic annotations based on unstructured text. These efforts are mainly focused on creating annotations in RDFa or Microdata format via editor interface. The approaches they adopt vary in terms of automation (e.g. usage of NLP techniques for named entity recognition and entity linking until some level). A comprehensive survey of semantic content authoring approaches and tools can be found in~\cite{Khalili20131}. 

The generation of semantic annotation based on dynamic data served in a structured way (e.g. relational databases, web APIs) is critical considering the volume of the data. This publication method mainly requires mappings from the metadata of the data source to a vocabulary such as schema.org. Triplify~\cite{Auer2009} uses an SQL-based lightweight mapping approach to create RDF out of relational data. D2RQ~\cite{Bizer04d2rq-} operates with a declarative mapping language and creates virtual RDF graphs on top of relational databases. RML~\cite{dimou_ldow_2014} provides a mapping language and a processor for mapping data from various sources including but not limited to XML files, web APIs, relational databases and CSV files to RDF. All of the aforementioned techniques can be used for creating JSON-LD since it is an RDF serialization format.

Our main contributions are a holistic methodology and a proof of concept for analyzing and mapping static and dynamic data to schema.org and create semantic annotations to enrich touristic service providers' content on the web. For the static data, we create the annotations manually, to ensure high accuracy and domain coverage. For the dynamic data, we create mappings based on the domain analysis and generate automated annotations externally, which makes the deployment to the web pages feasible, since it does not require a major software development effort on the touristic service providers' side.







\section{Methodology}
\label{sec:Methodology}
In this section, we describe our methodology to annotate the TVB Mayrhofen-Hippach website, the starting point of our effort to annotate touristic service providers in the region of Tyrol, Austria completely. 
Our methodology comprises three essential activities: (i) data sources and format analysis, (ii) information modeling, and
(iii) domain specification definition.

\subsection{Data sources and format analysis}
Information available on the TVB Mayrhofen-Hippach website are originating from external and internal sources, where the external data came mostly from Feratel\footnote{\url{http://www.feratel.at/}}.

\subsubsection{Feratel}
Feratel offers a destination management system\footnote{\url{http://www.feratel.at/en/solutions/feratel-destination/}} for services related to tourism and travel industry. The system provides information about accommodation, packages, events, etc., including a real-time  service to check room availability as well as to perform booking action.
The system is widely used by service providers such as hotels to manage their booking system, as well as by TVBs to market a region.
The system can be accessed through a web API so-called Deskline Standard Interface (DSI)~\cite{ebner2016} which serves data in an Extensible Markup Language (XML) format\footnote{\url{https://en.wikipedia.org/wiki/XML}}.

\subsubsection{TVB Itself}
The website also contains information which was created internally by the TVB itself by using a CMS.
At the moment this study was performed, the TVB uses TYPO3\footnote{\url{https://en.wikipedia.org/wiki/TYPO3}} to manage their self-created content, where information will be entered in the backend and will be presented in a semi-structured format. In this case, TVB has full access to the CMS, so they can manage and define their content structure including installing extensions.

\subsection{Information modeling}
After identifying the data sources and their formats as explained in the previous section, we start our next task to determine important concepts from every data source.
For data originating from Feratel, we consulted the APIs documentation, XML responses, as well as how the data were presented in the relevant webpages. For self-created data, we visited each relevant webpage and performed analysis on it.

We started the information modeling by identifying the types of web pages and what kind of information are presented there. We went through of each webpage and analyzed what kind of data can be annotated. For example, the webpage Die Region\footnote{\url{http://www.mayrhofen.at/die-region/}} contains the information about the Mayrhofen-Hippach Holiday Region, its villages, and the latest news and events from the region. The webpage of the Mayrhofen region\footnote{\url{http://www.mayrhofen.at/die-region/mayrhofen/}} contains the name of the region, url, location, information about the region, picture of it. This data are presented as text, image object, url. Another example is Globeseekers yoga event\footnote{\url{http://www.mayrhofen.at/en/events/detail/events/globeseekers-yoga/}}. The web page contains the name of the event, its location, dates and time, description, information on prices, organizer, contact information and image.

Further, we defined what primary categories and subcategories could be chosen from the menu, submenu and web content. We made the list of categories and subcategories and selected the most important. We selected seven main categories: TVB Mayrhofen, Mayrhofen Hippach region, Ski Areas, Accommodations, Infrastructure, Events, Articles. For example, Mayrhofen Hippach region is the category, which includes the subcategories with the information of Mayrhofen, Ramsau, Schwendau, Brandberg, Ginzling, Hippach regions. Event is the category for different types of events and activities in TVB Mayrhofen Hippach, such as concerts, lectures, conferences, festivals, etc. 

\subsection{Domain specification definition}
From the information modeling activities explained above, we obtained a list of identified information concepts (such as Place, News, Article, Event) including their attributes (such as name, location, start date, contact information). The next step will be to create a domain specification for every identified concept. 

In this step, we selected the right class from schema.org which most adequately describes a concept. We searched for suitable schema.org classes for every concept, defined a selection of these properties, selected the range types and recursively repeated this process when structured types appear as ranges. For example, according to our domain analysis for the type Hotel, we selected the following properties: name, description, telephone, faxNumber, email, url, currenciesAccepted, address, aggregateRating, geo, makesOffer, and image. Each property has its range, e.g. Text, Url, DateTime, QuantitativeValue. Some elements can have external properties, and we considered them too. Address in schema.org has the range PostalAddress, where PostalAddress is defined by addressCountry, addressLocality, addressRegion, postalCode and streetAddress. Property makesOffer has the range Offer with the following properties: name, availability, itemOffered, priceSpecification.
A subset of our domain specification is shown in Table~\ref{tab:specification}.

\begin{table}
\centering
\caption{A subset of our domain specification}
\label{tab:specification}
\begin{tabular}{clll}
\toprule
\textbf{No.} & \textbf{Type} & \textbf{Property} & \textbf{Range Type} \\
\midrule
1 & Hotel & & \\
  & & address & PostalAddress \\
  & & aggregateRating & AggregateRating \\
  & & currencyAccepted & Text \\
  & & description & Text \\
  & & geo & GeoCoordinates \\
  & & image & ImageObject \\
  & & makesOffer & Offer \\
  & & name & Text \\
  & & paymentAccepted & Text \\
  & & url & URL \\
2 & PostalAddress & & \\
  & & addressCountry & Text \\
  & & addressRegion & Text \\
  & & postalCode & Text \\
  & & ... \\
3 & AggregateRating & & \\
  & & ratingValue & Number \\
  & & reviewCount & Number \\
4 & GeoCoordinates & & \\
  & & latitude & Number \\
  & & longitude & Number \\
\bottomrule
\end{tabular}
\end{table}

\subsection{Discussion}
We would like to outline a few important things we encountered during our analysis, information modeling, and defining domain specification. First, we worked with two different types of data: the static data and dynamic data. Static data refers to rarely changed information, meaning that once created then the information will stay as the original. Fall into this category are the information about the region and TVB itself, ski areas, press releases, and articles. On the other hand, dynamic data are changed regularly, for example, the price of an offer for an accommodation room could frequently be changed, and therefore the annotation should be updated regularly as well. Second, when selecting a class in schema.org, we tried to be as specific as possible. For example, information about concert should be annotated with MusicEvent (more specific type) instead of Event (generic type). And therefore, in our domain specification, we also consider the super and subclasses relationships between types.

\section{Implementation}
\label{sec:Implementation}
In this section, we explain our implementation on how to annotate content of the TVB Mayrhofen-Hippach website.
Our implementation consists of a series of activities, starting from defining a mapping between data schema to classes and properties from the selected vocabulary.
Next, we use the mapping to perform annotation automatically or manually, and finally, we attach annotations to the target website.

\subsection{Automatic annotation of content}
\label{sub:automatic}

A significant part of the web content available on the TVB Mayrhofen-Hippach website is generated based on data made available by Feratel. 
Feratel provides information about events, accommodations, offers, and infrastructure for multiple regions in Austria including the Mayrhofen-Hippach region. We developed a software solution that automatically annotates events, accommodations, offers, and infrastructure in Mayrhofen-Hippach according to schema.org. 
Feratel data can be accessed through the DSI by accepting requests from a client, processes the request and produces responses, all in the format of XML according to a particular structure.
To annotate every event, accommodations, offer, and infrastructure item coming from Feratel we take the following approach:
\begin{enumerate}
\item Define a mapping between the Feratel data types to the specification produced in the domain specification definition explained in the previous section. 
\item Develop a software wrapper to communicate with the DSI, and consumes the mapping to produce annotations in a JavaScript Object Notation for Linked Data (JSON-LD)\footnote{\url{https://en.wikipedia.org/wiki/JSON-LD}} format.
\end{enumerate}

\begin{table}
\centering
\caption{Statistics of Data Types Mapping}
\label{tab:statistic}
\begin{tabular}{clrr}
\toprule
\multirow{2}{*}{\textbf{No}} & \multirow{2}{*}{\textbf{Description}} & \multicolumn{2}{c}{\textbf{Mapped Data Types}} \\ \cline{3-4}
 &  & \textbf{Feratel} & \textbf{Schema.org} \\
\midrule
1 & Accommodation & 19 & 6\\
2 & Event & 29 & 11\\
3 & Infrastructure & 212 & 67\\
\bottomrule
\end{tabular}
\end{table}

Table~\ref{tab:statistic} shows the statistics of the data types mapping, where not all types available in Feratel can be mapped into schema.org. For example, for infrastructure, we were able to identify 67 types available in schema.org from 212 types provided by Feratel.
There are various causes for the mapping deficits, such as: language differences, different conceptions and orientation for data types providing by Feratel and schema.org. In German language there exist a lot of names which represent the same type of object, for example Gasthof, Pension, Aparthotel, Berghotel, Berggasthof are hotels based on their characteristics and features. Feratel, as mentioned, offers data related to the tourism and travel industry. They categorized data based on types of objects in Austria. This classification is on the one hand too general (e.g. the type of events Theater/Show/Tanz/Film/Kleinkunst), and on another hand too detailed (e.g. the types Fahrrad-Transport, Fahrrad-Werkstätte, Fahrrad-Verleih, E-Bike. Schema.org provides the most common terms to annotate a great variety of entities on the web. But it is relatively young vocabulary and can't cover all content. That's why we had difficulty mapping some types from Feratel. For example, type SportsActivityLocation in schema.org contains only 9 subtypes, whereas in Feratel more than 40 are presented. Therefore for many types from Feratel we chose more generic classes from schema.org, such as: SportsActivityLocation, LocalBussiness, TouristAtraction, Store, Event, CivicStructure and LodgingBusiness.
Also, it is worth mentioning that in some cases, a data instance might inherit properties from multiples types, known as a multi-types entity.

\begin{figure}
\centering
	\includegraphics[width=\textwidth]{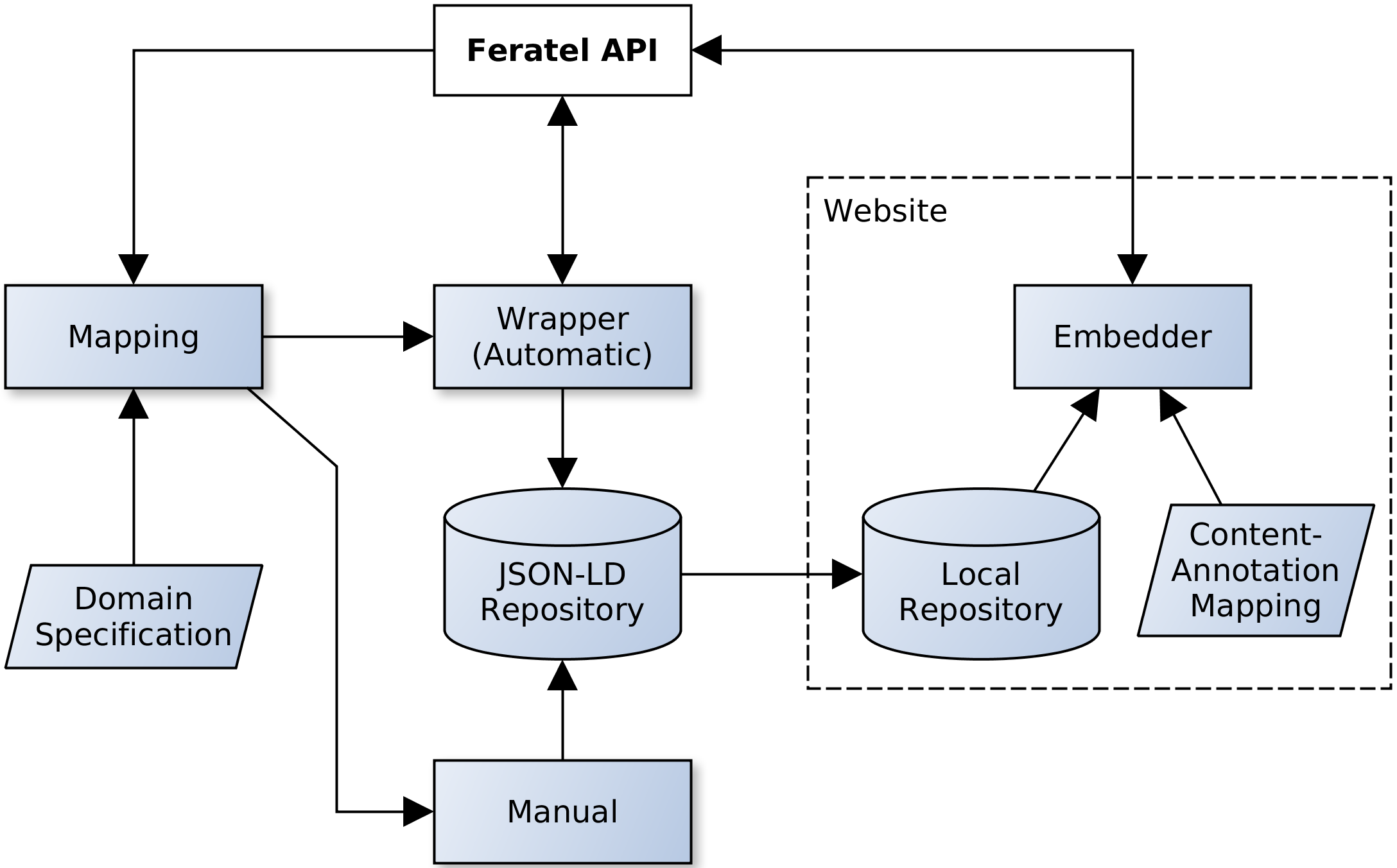}
\caption{Architecture of content annotation approach}
\label{fig:architecture}
\end{figure}

\begin{figure}
\centering
	\includegraphics[width=\textwidth]{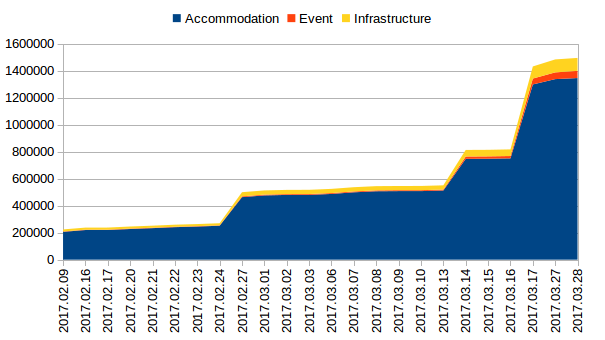}
\caption{The cumulative number of produced triples from automatic annotations}
\label{fig:daily-update}
\end{figure}

Figure~\ref{fig:architecture} shows the architecture of our approach for content annotations. For automatic annotation, the domain specifications will be aligned with the Feratel API specification to produce a mapping to be consumed by the wrapper.
When defining the mapping, we started with our previous work~\cite{Akbar2015}, where we tried to map each XML element and attribute of Feratel data to class and property in the specifications. 
A plugin installed in the website server loads annotation in JSON-LD format from a repository and embeds it into the associated webpage identified by a mapping (detail explanation in Section~\ref{sub:linking}).
The wrapper runs on a daily basis, producing incremental updates.
From about 1,6 million triples we produced per March 29, 2017, mostly dominated by information related to accommodation, as shown in Figure~\ref{fig:daily-update}. A full update needs to be performed whenever something changed in the domain specification, mapping, or wrapper implementation, e.g. when we supported multilingual annotation on 17.03.2017.

\subsection{Manual annotation of content}
\label{sub:manual}

Content not coming from content sources providing APIs, needs to be annotated manually. This is the case with ski areas, information about the Mayrhofen-Hippach region, press release articles, the TVB Mayrhofen-Hippach description and a few infrastructures which are not available in Feratel.

\begin{figure}
\centering
	\includegraphics[width=\textwidth]{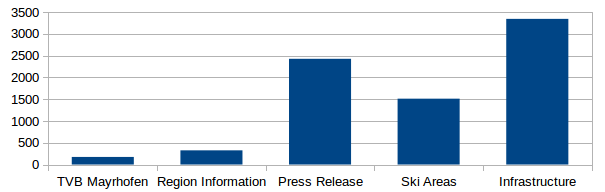}
\caption{Number of produced triples from manual annotations}
\label{fig:manual-update}
\end{figure}

A conceptual analysis was performed before creating manually the annotations. For infrastructure for example, as its content is more complex in structure as all the other types of content available on Mayrhofen website, a large set of concepts and properties had to be considered. The conceptual analysis task enabled us to identify all the relevant information available in the Mayrhofen-Hippach holiday region and then clearly define the structure~\cite{Ackstaller2016}.
As shown in Figure~\ref{fig:manual-update}, we were able to produce around 8 thousand triples for this manually produced annotations, mostly for infrastructures and press release articles.

We did not use any Natural Language Processing (NLP) technique for producing manual annotations due to time limitation. Finding the correct tuning parameters for an NLP algorithm requires data training which we do not have. Creating annotations manually was the best option, where an editor was used to guarantee annotations correctness and validity.

\subsection{Linking of content with annotation}
\label{sub:linking}
Once the annotations are created as described in the previous sections, what remains is to deploy them or in other words to link the annotations to the content which are available on TVB Mayrhofen website. This section describes our solution to achieve this goal. An important aspect is that the annotations and the content are available on different systems and are brought together via a deployment described in the rest of this section.  Two core requirements needed to be fulfilled by our solution:
\begin{enumerate}
\item How to connect the content (which is in HTML) residing on the website with its annotation (JSON files) available on another server
\item How to embed the annotation (JSON files) into the content (HTML)
\end{enumerate}

To fulfill the requirements, we designed our solution to separate the annotation process and embedding annotation to content process such that each process can be maintained without interfere with the other.
In Figure~\ref{fig:architecture}, the interlinking is done in the website server separately from the annotations processes.
The separated processes will be performed as follows:
\begin{enumerate}
\item Annotation process produces all required annotations from Feratel API through automatic annotations (Section~\ref{sub:automatic}) as well as from manually generated annotations (Section~\ref{sub:manual}). All annotations in JSON-LD format will be stored in a repository.
\item Embedder process which is installed as part of Content Management System (CMS) of the website loads a JSON-LD file from a local repository, where both repositories will be synchronized regularly. To identify which file should be loaded and embedded to a page, the Embedder reads a mapping between content and its associated annotation.
\end{enumerate}

\begin{table}
\centering
\caption{An example of content to annotation mapping}
\label{tab:mapping}
\begin{tabular}{|l|l|}
\hline
\textbf{Key} & \textbf{Value} \\
\hline
0000 & 0a2346a9-3b05-4dc4-a056-1f32ccf05fe8.json\\
\hline
1111 & 0a19990c-b879-42ff-acc1-886d1ea59365.json\\
\hline
/meta/impressum & impressum.json\\
\hline
/service/kontackt & contact.json\\
\hline
\end{tabular}
\end{table}

As shown in Figure~\ref{fig:architecture}, a plugin for Embedder will reside in the website (in this case in the CMS of TVB Mayrhofen website). When a request is received from a client, the plugin consults its mapping database, if a matching is found, then it will load the file from the local JSON-LD files repository and embed its content into the HTML response to the client.
Items of the mapping database in the format of \texttt{<key, value>}, represent an association between a webpage and its annotation.
In the current implementation, we have two types of association mapping: we use \texttt{<Page-ID, Feratel-ID.json>} and \texttt{<Page-URL, Filename.json>} for the data coming from Feratel and manually annotated respectively.
``Page-ID'' and ``Page-URL'' were obtained from the CMS and ``Feratel-ID'' from the Feratel API.
A small fragment of the mapping is shown in Table~\ref{tab:mapping}, where an annotation will be identified either with an identification number for Feratel data and an URL for annotation generated from information on the website.

\section{Result and Evaluation}
\label{sec:Result}
In this section, we list and discuss the results of our work. After that, we explain the results of our qualitative and quantitative evaluations.

\subsection{Results}
As results, we were able to annotate numerous topics of information from the TVB Mayrhofen-Hippach website:
\begin{enumerate}
\item \textbf{Accommodation}, information related to accommodation including offered places such as a room that can be rented to stay for a given period.
\item \textbf{Event}, information about events that are happening at a particular time and location.
\item \textbf{Infrastructure}, information which is related to physical businesses or organization, including places that someone may find interesting (point of interest).
\item \textbf{Organization}, information which is related to the TVB itself, for example, its address and contact point, opening hours, etc.
\item \textbf{Press-release}, information which is related to news article or report including blog posting.
\item \textbf{Region}, information which is related to the region of Mayrhofen Hippach, for example, content about Mayrhofen and its villages, family holiday guides, winter guides, etc.
\item \textbf{Ski-area}, information which is related to ski-area such as ski-resort, ski-lift or slope which currently receives a minimal support in schema.org.
\end{enumerate}

\subsection{Evaluation}


\begin{figure}
\centering
	\includegraphics[width=.9\textwidth]{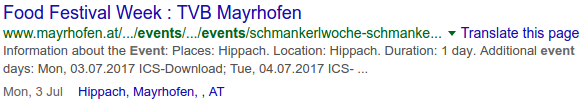}
\caption{Google Search rich-snippet for an event}
\label{fig:rich-snippet}
\end{figure}

\begin{figure}
\centering
	\includegraphics[width=.55\textwidth]{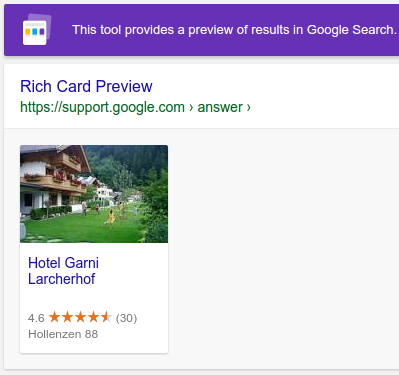}
\caption{Google rich-card preview for a hotel}
\label{fig:rich-card}
\end{figure}

For qualitative evaluations, we monitored Google's search engine results especially the appearances of rich results.
As shown in Figure~\ref{fig:rich-snippet}, Google Search was able to show events in a more structured way where the date of an event, as well as its location, will be additionally included in the search results.
Google also produces richer and structured information for a hotel as shown in Figure~\ref{fig:rich-card}, where the rating and location for the hotel will be included in a rich card.

\begin{figure}
\centering
	\includegraphics[width=.95\textwidth]{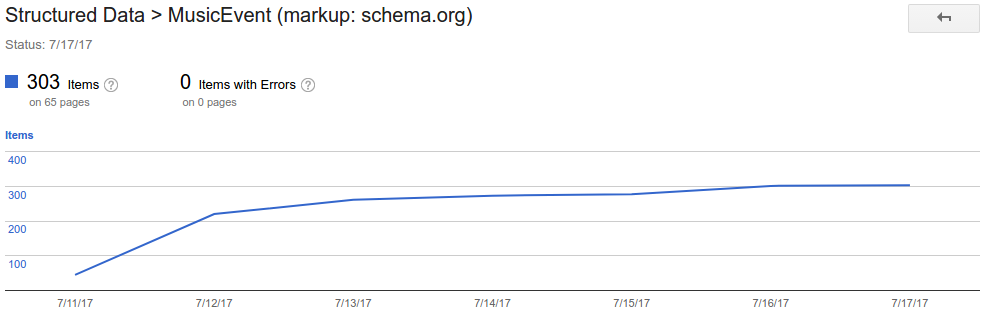}
\caption{Google search console for structured data of MusicEvent}
\label{fig:search-console}
\end{figure}

For quantitative evaluations, we use the Google's Search Console\footnote{\url{https://www.google.com/webmasters/tools/}} to measure a few aspects:
(i) how long was required to detect the annotated pages,
(ii) how often the annotated pages were crawled,
(ii) how many errors were detected.
As an example, we took the statistics of structured data of ``MusicEvent'' as shown in Figure~\ref{fig:search-console}. From more than 300 detected items, 2 days were required for collecting about 220 items (status on 7/13/2017, compared to the status on 7/11/2017).
Further, we used the ``Last.Checked'' history of Google's search console to measure how often the annotated pages were crawled.
From all detected structured data of ``LodgingBusiness'', which were crawled for more than 50 times (status on 7/17/2017), the average of crawled frequency was 2 days. And lastly, there was no error detected.

\section{Ongoing work and outlook}
\label{sec:OngoingWork}
Besides the work described in this paper, we consider several more directions to continue and extend this topic. On the way to provide a holistic armamentarium for a semantic web contribution for touristic service providers, our ongoing and future work comprises of the tasks and ideas described below.

\subsection{Schema.org 3.1 and actions}
\label{sec:sdo31}
Since May 2015 schema.org offers a new extension mechanism that facilitates the creation of specialized and/or deeper vocabularies based on the core vocabulary of schema.org. We have submitted an accommodation extension to schema.org that became an integrated part of schema.org 3.1 \cite{karle2017extending}. As future work we will update the set of annotations produced so far to be fully aligned and complete regarding with schema.org 3.1, particularly the hotel extension. 
Once accommodation offers and booking data are machine readable, a system that makes use of these data to enable automatic direct booking of offers can be established. 
The overall approach includes three main steps: 1) annotation of booking data, including data about room, offers, etc. This kind of data needs to be annotated with schema.org including the schema.org hotels extension as part of schema.org 3.1, 2) annotation of booking engines, meaning booking engines need to be annotated with schema.org in order to be found and understood and 3) implementation of an automated direct booking agent. As part of this last step a booking agent connects the booking data with the booking engine, crawls the booking data and the booking engines/endpoints and performs heuristic reasoning due to the fact that the booking data is usually not complete, approximately, and partially inaccurate. Our work so far has focused on the first step. As part of the current and future work we are tackling step 2 and 3 providing semantic annotations using schema.org actions and then designing and implementing an automated direct booking agent.

\subsection{Chat bots and Intelligent Personal Assistants (IPAs)}
\label{sec:chatbots}
Since the introduction of chat bots on Facebook's F8 conference in April 2016, the topic has attracted a lot of attention from small and large companies alike. Big names in the software industry, including Amazon, Apple, Facebook, Google and Microsoft are developing their own solutions and are opening their APIs providing support for developers to build chat bots and personal assistants for their platforms. 
Tourism is a domain where chatbots and intelligent assistants have an immediate applicability from finding touristic service providers, their services and offers to booking/buying these items using new conversational, intuitive interfaces. As current work we are developing a chatbot for Mayrhofen-Hippach region. The bot is named as ``Mayley'', available as a Facebook Messenger bot as well inside a web widget deployable target website\footnote{\url{https://www.facebook.com/MayleyMayrhofen}}. 
Mayley uses directly the Mayrhofen-Hippach region semantic annotations to create the appropriate answers given tourists natural language requests. Combined with user profile information and rules, Mayley delivers personalized content to its users. We are also working on using semantic annotations to update the set of entities and intents the chatbot understands.

\subsection{Schema.org annotation generation platform}
\label{sec:generator}
The generation of schema.org annotations for web content is, especially for people with no background in programming languages or semantic technologies, not trivial. But often these are the people who should actually make use of schema.org annotations: content creators for enterprises or touristic service providers, event promoters or blog editors. Various different factors keep them from annotating their content on the web, broken down, three main challenges emerge: (1) what vocabulary to use, (2) how to create JSON-LD files and (3) how to import annotation files into websites. To tackle those three challenges we are working on a web platform which offers assistance in the whole annotation creation process. 
This platform should remove a big obstacle between the "normal" web content creator and the semantic web.

\subsection{Validation of schema.org annotated data}
\label{sec:validator}
A study about the use of schema.org in the hotel domain\cite{karle2016there} showed, that a lot of content creators, enterprises and touristic service providers on the web want to use semantic annotations on their website, but are not able to do it in a correct way. So besides providing a solution to create and publish semantic annotations (mentioned in Section \ref{sec:generator}) we also work an a means to validate existing semantic annotations. There are several validators around on the web, with Google's Structured Data Testing Tool\footnote{\url{https://search.google.com/structured-data/testing-tool}} leading the way, but those validators either only validate for syntactical correctness or do only very limited or biased semantic validation. The solution we work on is based on two different fundamentals. First of all the recommended or required vocabulary is defined. 
The second foundation is a set of rules to define correlation between schema.org properties and their ranges as well as the correlation between different properties. 
More information about the work on that idea can be found in \cite{simsek2017domain}. This tool allows content creators on the web to not only generate some mandatory "meta-tags" but valid, high quality annotations which lead to reusable, high quality web content.

\subsection{Touristic knowledge graph}
\label{sec:knowledgegraph}
With the idea of a touristic knowledge graph, we want to support tourists as well as touristic service providers and also provide a means for analyzing touristic developments over time. The knowledge graph comprises information about touristic services, the infrastructure of a region, points of interests but also information about arrivals of guests, events, whether data and other factors influencing tourism. This idea is in a very early phase and will be followed in the futures.

\section{Conclusion}
\label{sec:Conclusion}
In this paper, we presented the work done to provide better online visibility for touristic service providers by using semantic technologies, particularly semantic annotations using schema.org. We used Mayrhofen-Hippach region as a pilot, created and deployed for the touristic services providers in this region a substantial amount of annotations.
The annotations covered a wide variety of information topics including events, accommodations and accommodation offers, ski areas, the region, press release articles, the organization itself and a large variety of infrastructure information. Moreover, most of the annotations need to be regularly updated on a daily basis due to the dynamic changes in the data sources, for example, the price of an accommodation offer. 
The annotations were created in a mixed manner, automatically and manually, where the software tools for generating annotations automatically as well as tools to support the human users to create annotations are currently deployed for internal use only. After a few improvements, including integration with latest recent mapping languages such as RML, we will offer them as services to be used not only by TVBs but also by any organization that willing to annotate their webpages.

The same approach has been successfully applied to other TVBs, namely Seefeld\footnote{\url{https://www.seefeld.com/}} and F{\"u}gen\footnote{\url{https://www.best-of-zillertal.at/}} among others.
We also applied the approach to some other use cases in the tourism industry, including ski schools, ski resorts, golf places.
And currently, we are working to annotate an interactive map provided by General Solutions\footnote{\url{https://general-solutions.eu/}}. The map contains rich geo-related information such as hiking or biking routes, entry points for a route as well as point of interests along a route.
Our ultimate goal is to be able to annotate all tourism relevant information in the region of Tyrol, Austria, not only to increase the online visibility of the region but also to enable intelligent applications to run on top of them.

\section*{Acknowledgements}
This work was partially supported by the EU project EUTravel. We would like to thank Daniel Ackstaller, Daniel Eppacher, Christian Esswein, Omar Holzknecht, Philipp Kratzer, Jonas Stock, Johannes Strickner, Simon Targa, Sahin Ucar, Hannes Vieider, and Jakob Winder for their fruitful discussions and input.

\bibliographystyle{splncs03}
\bibliography{chapters/bib}

\end{document}